\documentstyle[12pt]{article}
\begin{document}

\title{A new class of (3+1) quantum geometric models} 

\author{Ion I. Cot\u aescu\\ The West University of Timi\c soara,\\
        V. P\^ arvan Ave. 4, RO-1900 Timi\c soara, Romania}

\date{\today}

\maketitle

\begin{abstract}
A new family of analytically solvable quantum geometric  models  is 
proposed. The structure of the energy spectra as well as the form of the 
corresponding eigenfunctions are presented pointing out their main specific 
properties.  
  
\end{abstract}
\
\

In the general relativity  the relativistic  (three-dimensional isotropic) 
harmonic oscillator has been defined as a free system on the  anti-de Sitter  
static background \cite{M}. In a previous article we have generalized this 
model  to a family of models the metrics of which represent suitable 
deformations of an anti-de Sitter static metric \cite{P}. These are the 
spherically symmetric static metrics given by the line elements which ca be 
written  in spherical coordinates and natural units ($\hbar=c=1$) as 
\cite{COTA}
\begin{equation}\label{(m0)}
ds^{2}=\frac{\alpha}{\beta}dt^{2} -
\frac{\alpha}{\beta^{2}}dr^{2} -
\frac{r^{2}}{\beta}(d\theta^{2}+\sin^{2}\theta d\phi^{2}),
\end{equation}
where the functions 
\begin{equation}\label{(ab)}
\alpha=1+(1+\lambda)\omega^{2}r^{2}, \qquad 
\beta=1+\lambda\omega^{2}r^{2} 
\end{equation}
depend on the real parameter $\lambda$. We have shown \cite{COTA} that the 
quantum models of the free massive scalar particles on these backgrounds are of 
two kinds. The first set contains the models with $\lambda<0$. The particle  
of a such a model is confined to a spherical cavity having a countable energy 
spectrum with a fine-structure due to a rotator-like term. In the 
particular case of $\lambda=-1$, when the  metric is just the anti-de Sitter 
one, the spectrum becomes equidistant and the fine-structure disappears 
\cite{M}. The other set, corresponding to  $\lambda>0$, is of  models 
with mixed spectra, having  finite discrete sequences with fine-structure and 
continuous parts. The model  with $\lambda=0$ is of a special interest  since 
it is very closed to the nonrelativistic isotropic harmonic oscillator which, 
although, is just the nonrelativistic limit of all these models.  Another  
characteristic  of this family is that  the quantum numbers which determine the 
discrete energy levels, the main quantum number and the angular momentum one,  
are separately involved  in two different terms of the level formulas. For this 
reason these models, except the anti-de Sitter one, have been called 
relativistic rotating oscillators.

These results indicate that  new  families of analytically solvable geometric 
models with more sophisticated behavior could exist. By looking for them, we 
have found another interesting one we would like to present in the following. 
This new family has the metrics given by   
\begin{equation}\label{(m)}
ds^{2}=\frac{\alpha}{\beta}dt^{2}-\frac{\alpha}{\beta^{2}}dr^{2} -
r^{2}(d\theta^{2}+\sin^{2}\theta d\phi^{2})
\end{equation}
where the functions $\alpha$ and $\beta$ are just those defined by (\ref{(ab)}) 
depending on the real parameter $\lambda$. We note that the  model with 
$\lambda=0$ is the same as that of the family defined by (\ref{(m0)}). Moreover,  
the general behavior with respect to $\lambda$ of  both these families is 
similar. Thus, we observe that when $\lambda<0$ the event horizon  of an 
observer situated at $x^{i}=0$ is the sphere of the radius $r=r_{e}=1/\omega 
\sqrt{-\lambda}$ where the metric is singular.  For non-negative $\lambda$ this 
is at $r_{e}=\infty$. Here we shall consider  the free motion only 
on the domain $D$ of the space coordinates bounded by the event horizon, i.e. 
$r\in  [0,r_{e})$. Obviously, since  all these metrics are invariant under time  
translations and  space rotations, the energy and the angular momentum of the 
free motion are conserved.

Let  $\phi$ be a scalar massive field of the mass $M$, defined on $D$, 
minimally coupled with the gravitational field \cite{B1}. Because of the energy 
conservation, the Klein-Gordon equation
\begin{equation}\label{(kg)}
\frac{1}{\sqrt{-g}}\partial_{\mu}\left(\sqrt{-g}g^{\mu\nu}\partial_{\nu}\phi
\right) + M^{2}\phi=0,
\end{equation}
where $g=\det(g_{\mu\nu})$, admits the fundamental solutions 
\begin{equation}\label{(sol)}
\phi_{E}^{(+)}(x)=\frac{1}{\sqrt{2E}}e^{-iEt}U(r,\theta,\phi), 
\quad \phi^{(-)}=(\phi^{(+)})^{*},
\end{equation}
which must be orthogonal with respect to the relativistic scalar product 
\cite{B1} 
\begin{equation}\label{(sp1)}
<\phi,\phi'>=i\int_{D}d^{3}x\sqrt{-g}g^{00}\phi^{*}(x)
\stackrel{\leftrightarrow}{\partial_{0}} \phi'(x).
\end{equation}
The conservation of the angular momentum allows one to separate the variables 
of the Klein-Gordon equation (\ref{(kg)}) by choosing particular solution 
(\ref{(sol)}) with   
\begin{equation}\label{(u)}
U(r,\theta,\phi)=R_{E,l}(r)Y_{lm}(\theta, \phi),
\end{equation}
where $Y_{lm}$ are the usual spherical harmonics and $R_{E,l}$ are the radial 
wave functions which satisfy the radial equation
\begin{equation}\label{(kg1)}
-\beta\frac{d^{2}R_{E,l}}{dr^{2}}-(3\beta -1)\frac{1}{r} \frac{dR_{E,l}}{dr}+
\frac{\alpha}{r^{2}\beta}l(l+1)R_{E,l}-\left(E^{2}-
\frac{\alpha}{\beta} M^{2}\right)R_{E,l}=0.
\end{equation}
Moreover, from (\ref{(sp1)}) it results that the scalar product reduces to  the 
radial one,  
\begin{equation}\label{(psc)}
<R,R'>=\int_{0}^{r_{e}}\frac{r^{2}dr}{(1+\lambda\omega^{2}r^{2})^{1/2}}R^{*}(r)
R'(r).
\end{equation}

Now we have all the elements for deriving  the energy spectra and  the 
corresponding energy eigenfunctions. All the results in the  case of 
$\lambda=0$, which might be separately treated, are known from Ref.\cite{COTA}.   
Therefore, we can start directly with the general case of any $\lambda\not=0$ 
where it is convenient to use the new variable $y=-\lambda\omega^{2}r^{2}$, 
and the notations
\begin{equation}\label{(nu)}
\nu=\frac{1}{4\lambda\omega^2}\left[\lambda\omega^{2}+\left(1+
\frac{1}{\lambda}\right)M^{2}-E^{2} \right].
\end{equation}
We shall look for a solution of the form
\begin{equation}\label{(100)}
R(y)=N(1-y)^{k/2}y^{l/2}F(y),
\end{equation}
where $k$ is a real number and $N$ is the normalization factor. After 
a few manipulation we find that, for  
\begin{equation}\label{(pau)}
k^{2}-k-\frac{M^2}{\lambda^{2}\omega^2}+\frac{1}{\lambda}l(l+1)=0,
\end{equation}
the equation (\ref{(kg1)}) transforms into the following hypergeometric 
equation
\begin{equation}
y(1-y)\frac{d^{2}F}{dy^2}+\left[l+\frac{3}{2}-y(2p+1)\right]\frac{dF}{dy}-
(p^{2}-\nu)F=0.
\end{equation}
where we have denoted
\begin{equation}
p=\frac{1}{2}(k+l+1)
\end{equation}
This equation has the solution \cite{B2}
\begin{equation}\label{(F1)}
F=F(p-\sqrt{\nu}, p+\sqrt{\nu}, l+\frac{3}{2}, y),
\end{equation}
which depends on the possible values of the parameter $p$. From 
(\ref{(pau)}) it follows that  
\begin{equation}\label{(kpm)}
 k=k^{\pm}_{l}=\frac{1}{2}\left[ 1\pm \sqrt{1+ 
4\left(\frac{M^2}{\lambda^{2}\omega^{2}}-\frac{1}{\lambda}l(l+1)
\right)} \right].
\end{equation}
which means that we have two possibilities, namely
\begin{equation}
p=p^{\pm}_{l}=\frac{1}{2}(k^{\pm}_{l}+l+1)
\end{equation}

Furthermore, we observe that for 
\begin{equation}\label{(quant1)}
\nu=(p+n_{r})^{2}, \quad n_{r}=0,1,2...,
\end{equation}
$F$ reduces to a polynomial of degree $n_{r}$ in $y$.  
According to  these results, we can establish the general form of the 
solutions of (\ref{(kg1)}), which could be square integrable with respect to 
(\ref{(psc)}). This is 
\begin{equation}\label{(u1)}
R_{n_{r},l}(r)=N_{n_{r},l}(1+\lambda\omega^{2}r^{2})^{k/2}r^{l}F(-n_{r},
2p+n_{r}, l+\frac{3}{2}, -\lambda\omega^{2}r^{2}).
\end{equation}
The radial quantum number, $n_{r}$, and $l$ can be embedded into   the main 
quantum number $n=2n_{r}+l$. It is obvious that  $l$ will take the even 
values from $0$ to $n$ if $n$ is even and the odd values from $1$ to $n$ for 
each odd $n$. 
Now by using (\ref{(quant1)}), (\ref{(pau)}) and (\ref{(nu)}) we find the 
general form of the quantization condition
\begin{equation}\label{(el)}
{E_{n,l}}^{2}-M^{2}\left(1+\frac{1}{\lambda}\right)
=\lambda\omega^{2}[1-(n+k+1)^{2}]. 
\end{equation}  
which involves the both quantum numbers, $n$ and $l$, since $k$ depends on $l$ 
as it results from (\ref{(kpm)}). How may be chosen its concrete value  we 
shall see in the following.

Let us first take  $\lambda>0$. In this case $r_{e}=\infty$, and the 
solution (\ref{(u1)}) will be  square integrable only if 
the condition $n+k+1<0$ is accomplished. This means that we must take 
$k=k^{-}_{l}<0$ (and $p=p^{-}_{l}$). 
Therefore, the discrete energy spectrum will have a finite number of levels  
with $n$ and $l$ selected by the pair of conditions 
\begin{equation}
n+k^{-}_{l}+1<0, \qquad l\le n.
\end{equation}
From (\ref{(el)}) it results that all these levels satisfy
\begin{equation}
E_{n,l}<E_{lim}= 
\sqrt{\lambda\omega^{2}+
M^{2}\left(1+\frac{1}{\lambda}\right)}.
\end{equation}
On the other hand, from (\ref{(nu)}) we observe that,  for  $E\ge E_{lim}$, 
$\nu$ is  negative or zero and then the hypergeometric functions (\ref{(F1)}) 
cannot be reduced to polynomials but remain analytic for negative arguments. 
Therefore, the functions 
\begin{equation}
R_{\nu,l}(r)=N_{\nu,s}(1+\lambda\omega^{2}r^{2})^{k^{-}_{l}/2}r^{l}F(p^{-}_{l}
-\sqrt{\nu},
p^{-}_{l}+\sqrt{\nu},l+\frac{3}{2},-\lambda\omega^{2}r^{2})
\end{equation} 
can be interpreted as the non-square integrable solutions corresponding to 
the  continuous energy spectrum, $[E_{lim}, \infty)$.

In the case of $\lambda<0$ the radial domain is finite the particle being 
confined to the spherical cavity of the radius $r_{e}=1/\omega \sqrt{-\lambda}$.  
The polynomial solutions (\ref{(u1)}) will be square integrable over 
$[0,r_{e})$ only if they are regular at $r_{e}$. 
This require the choice  $k=k^{+}_{l}$  (and $p=p^{+}_{l}$) for which there 
are no restrictions on 
the range of $n$. Therefore, the discrete spectrum is countable. Moreover, in 
this case we have no continuous spectrum since the hypergeometric functions 
(\ref{(F1)}) generally diverge for $y\rightarrow 1$ (when $r\rightarrow r_{e}$).    
It is interesting to note that for the particular values of the parameters 
for which we have
\begin{equation}
M^{2}\left(1+\frac{1}{\lambda}\right)
=-\lambda\omega^{2}
\end{equation}
the energy levels become  
\begin{equation}
E_{n}=\sqrt{|\lambda|}\omega(k^{+}_{l}+n+1),
\end{equation} 
linearly depending  on $n$ but keeping their fine-structure.

Thus we have found the energy spectra and the energy eigenfunctions up to the  
normalization constants. These show that the models we have studied can not be 
considered as rotating oscillators because of the energy term involving 
simultaneously both the quantum numbers $n$ and $l$. Moreover, the space 
behavior is also different since here the parameter $k$ of (\ref{(u1)}) depend 
on the quantum number $l$ while in the case of the rotating oscillators this 
was a constant independent on $l$.

However, despite of these differences, there are some similar features. We 
refer  especially to  the continuity with respect to $\lambda$ in 
$\lambda=0$ and to the nonrelativistic limit. Indeed,  we can prove that  the 
solutions we have obtained  are continuous with respect to $\lambda$, as in the 
case of the rotating oscillators \cite{COTA}.  Based on this property we can 
calculate the nonrelativistic limit  by taking $\lambda \rightarrow 0$ and, 
in addition, $M\gg \omega$ (i.e. $Mc^{2}\gg \hbar\omega$ in usual units).  The 
conclusion is that all the models  we have studied here have the same 
nonrelativistic limit  as that of the rotating oscillators, namely  the usual 
isotropic harmonic oscillator.

\end{document}